\documentclass[twocolumn,amsmath,amssymb,prb,longbibliography]
{revtex4-2}
\usepackage{xcolor}
\usepackage{multirow}
\usepackage{array}
\usepackage{amsmath}
\usepackage{graphicx}
\usepackage{dcolumn}
\usepackage{bm}
\usepackage{verbatim}
\DeclareMathAlphabet \mathbfcal{OMS}{cmsy}{b}{n}

\begin{document}

\title{Spin Configurations of Anyonic excitations in Moir\'e Fractional Chern Insulators }

\author{ Vadym Apalkov$^a$}
\author{ Tapash Chakraborty$^b$}

\affiliation{$^a$Department of Physics and Astronomy, Georgia State University, Atlanta, Georgia,30303, USA, $^b$ Department of Physics and Astronomy, University of Manitoba, Winnipeg, MB, Canada}

\date{\today}

\begin{abstract}
We have examined the nature of elementary excitations in particular, their spin polarization, in recently discovered zero-field
fractional Chern insulator in twisted bilayer MoTe$_2$. We have found that
the finite-size bilayer system, in close proximity to a planar parabolic quantum dot containing a single
electron (or a hole), forms an incompressible spin-unpolarized fractional Chern insulator at 1/3 filling
factor. The corresponding energy gap is found to be large and weakly dependent on the separation between the  quantum dot
and the bilayer. We predict that in this fractional Chern insulator coupled to a quantum dot,  spin-unpolarized quasiparticles and quasiholes are expected to appear as low-energy elementary excitations. The charge density results indicate that, for the spin-unpolarized quasiparticles, the positive charge is strongly localized near the negatively charged QD within the spin-flipped component, whereas for the spin-polarized quasiparticles, such localization is much less pronounced.
\end{abstract}

\maketitle

Discovery of the quantum Hall effects, more than four decades ago had a transformative
effect in the field of condensed matter physics\cite{40_years_QHE,
Two_Dimensional_Magnetotransport,
Two_dimensional_electron_correlation,Anomalous_Quantum_Hall_Effect,
Book_FQHE}. Originated in a semiconductor system forming the two-dimensional (2D) electron gas subjected to a strong perpendicular magnetic field, the quantum Hall effects influenced a vast range of physics outside the condensed matter, e.g., the cold-atom systems, and thanks to the fortuitous connection of the Hall conductivity to a topological invariant, a multitude of topological systems. In addition to its fundamental significance, the integer effect has also provided an important application as the resistance standard, and the von Klitzing constant\cite{Klitzing_encyclopedia}. Its understanding involves Landau quantization and the gap between the Landau levels.

Discovery of the fractional quantum Hall effect (FQHE), in particular, was explained by 
Laughlin\cite{Anomalous_Quantum_Hall_Effect,
Book_FQHE} as a consequence of strong electron-electron interactions that leads to a
uniform density liquid state -- the Laughlin state. That incompressible state is associated with
fractionally-charged low-energy quasiparticle-quasihole excitations and obey fractional
statistics\cite{Encyclopedia,Anyon_collision,2d_anyon_gas}. FQHE was found to be very versatile: in addition to its discovery in 
semiconductor heterostructure, this incompressible state  also appears in
 monolayer graphene\cite{FQHE_graphene}, bilayer graphene\cite{phase_transitions_bilayer,stable_pfaffian_bilayer}, trilayer graphene\cite{Apalkov_trilayer_PRB_2012}
and much else besides.

One essential ingradient for the effects to occur is a strong, perpendicular, externally
applied magnetic field. The celebrated Laughlin wave function for the ground state
 was composed for fully spin-polarized electrons, precisely because of the presence of this strong 
magnetic field. Given the profound understanding the FQHE has brought in the field of
many-body condensed matter, it sounds almost irreverent  to question the role
of the magnetic field (and the resulting Landau levels) for observation of the effect.
Interestingly, in two groundbreaking publications\cite{Fractional_QH_zero_magnetic_2011,
FQHE_absensse_Landau_2011}
 the authors did just that. These authors demonstrated that the fractional quantum Hall-like physics
can occur in {\it lattice} systems (unlike the isotropic, uniform-density fluid state of 
Laughlin) without the external magnetic fields.

Two-dimensional (2D) semiconducting moir\'e materials have recently
emerged as a suitable platform for exploring novel quantum phenomena.
As an example, rotating two planar sheets of MoTe$_2$ at small twist angles
$(~2^\circ - 4^\circ)$ relative to each other creates an interference pattern 
whose wavelength far exceeds the atomic spacing\cite{Berashevich_PRB_2011,my_optical_transition_twisted_bilayer_PRB_2011}. 
In these moir\'e
materials, flat and isolated Chern minibands are the lattice analogs of the
Landau levels that give rise to the new version of the FQHE initiated in \cite{Fractional_QH_zero_magnetic_2011,
FQHE_absensse_Landau_2011}. In this system the narrow
bandwidth quenches the kinetic energy of the electrons, thereby allowing 
the electron correlations to dominate. A partially filled Chern band allows
for the appearance of the fractional Chern insulator (FCI)
\cite{Signature_FCI_in_TMDC_2023,Optical_signature_1_3_TMDC_2026}.
In fact, experimental confirmation of the existence of the FCI have led to an explosive
growth of the field\cite{Recent_development_Chern_Insulator_Bergholtz_2024,
FCI_PRL_2011}. In addition to the degenerate ground states at fractional filling
factors -- the core tenets of the FQHE in a periodic 
geometry\cite{Ground_state_1_3_Halperin_1983}, the fractionally charged 
quasiparticles and quasiholes are also expected here (just as in Laughlin's theory) as elementary 
excitations\cite{Anomalous_Quantum_Hall_Effect,
Book_FQHE,
Chakraborty_PRB_1985}, in a zero external magnetic field.

Here we focus on the nature of low-energy elementary excitations, in particular their spin properties. The electron
spin played no significant role in the earliest understanding of the FQHE, where a large 
Zeeman energy was enough to create only fully spin-polarized quantum Hall states. However,
it was realized soon after that electron-electron interactions could actually lead to quantum Hall states
with non-trivial spin configurations\cite{Electron_spin_transitions_Charkraborty_2000,
Halperin1983TheoryOT,Role_of_reversed_spins_Tapash_1984,
ELementary_excitatons_spin_Tapash_1986,
Spin_reversed_Tapash_1986,
Spin_reversed_ground_state_Tapash_1990,
Thermodynamics_spin_polarized_Tapash_1996,
Half_polarized_PRL_2001,
Temperature_dependence_spin_polarization_Tapash_1999,
Spin_configurations_FQHE,
Evidence_phase_transition_FQHE,
FQHE_Maksym_World_1989,
spin_dependent_Maksym_1989}. In the case of the FCI, in the absence of Zeeman energies, 
spin-reversed quasiparticles\cite{ELementary_excitatons_spin_Tapash_1986,
Spin_reversed_Tapash_1986} are expected to be energetically favorable\cite{Spin_excitations_FCI_2026}.
We predict here that the bilayer FCI system, when Coulomb coupled to a planar, parabolic quantum dot containing an electron or a hole, facilitates the formation of quasiparticles and quasiholes with a large gap, that are spin-reversed.

{\it Theoretical approach:} As mentioned above, we consider a twisted bilayer MoTe$_2$ system. The corresponding moir\'e pattern is characterized by a twist angle $\theta $. For a small twist angle, the moir\'e pattern has a periodic structure with the lattice constant of $a_M = a_0/\theta $. Below, we consider the twist angle of $3.89^0$, for which fractional Chern insulators were predicted in numerical simulations
\cite{PRL_2023_FCI_MacDonald,FCI_Reddy_PRB_2023,
FCI_PRL_2024,FCI_first_principle_continuum_2024}. Here $a_0 = 3.52$ \AA  \, is the lattice constant of a MoTe$_2$ monolayer. The primitive lattice vectors of the moir\'e structure are defined as 
\begin{eqnarray}
& & \vec{a}_1 = a_M \left( \frac{\sqrt{3}}{2}, -\frac{1}{2}\right) \\
& & \vec{a}_2 = a_M \left( \frac{\sqrt{3}}{2}, \frac{1}{2}\right) ,
\end{eqnarray}
and the corresponding reciprocal vectors has the form
\begin{eqnarray}
& & \vec{b}_1 = b_M \left( \frac{1}{2}, -\frac{\sqrt{3}}{2}\right) \\
& & \vec{b}_2 = b_M \left( \frac{1}{2}, \frac{\sqrt{3}}{2}\right) .
\end{eqnarray}
where $b_M = \frac{4\pi }{\sqrt{3}  a_M}$. The vectors $\vec{b}_1$ and $\vec{b}_2$ define a reciprocal unit cell. 

We consider a AA-stacked bilayer where the single-electron low-energy states are characterized by a strong correlation between the spin and valley degrees of freedom. Specifically, in the coupled $K$ valleys of the two layers, the electronic states are spin-up, whereas in the coupled $K^{\prime }$ valleys, they are spin-down. Consequently, the low-energy effective Hamiltonian of the Transition Metal Dichalcogenides (TMDC) bilayer for a spin-up electron takes the form\cite{PRL_2023_FCI_MacDonald}
\begin{equation}
{\cal H}_{\uparrow} = \left(  
\begin{array}{cc}
- \frac{\hbar ^2 }{2m^*} \left( \vec{k} -\vec{\kappa}_{+} \right)^2 + \Delta_b  & \Delta_T \\
\Delta_T^*   &  - \frac{\hbar ^2 }{2m^*} \left( \vec{k} -\vec{\kappa}_{-} \right)^2 + \Delta_t 
\end{array}
\right) \label{Ham0}
\end{equation}
where $\vec{k}$ belongs to the reciprocal unit cell, 
$\vec{\kappa}_{+} = \frac{b_0}{2} \left( 1, \frac{1}{\sqrt{3}} \right) $ and 
$\vec{\kappa}_{-} = \frac{b_0}{2} \left( 1, -\frac{1}{\sqrt{3}} \right) $ are the positions of the $K$ points of the top and the bottom layers, respectively. The effective mass is $m^*=0.62 m_0$\cite{PRL_2023_FCI_MacDonald}, where $m_0$ is the electron mass. The Hamiltonian of the spin-down electron system is obtained by applying the time-reversal operator to the Hamiltonian ${\cal H}_{\uparrow}$. The Hamiltonian (\ref{Ham0}) describes the low-energy electron state in the valence band.

Potentials $\Delta_b$, $\Delta_t$, and $\Delta_T$ depend on the lattice vectors and couple the states with the wave vector $\vec{k}$ to the higher-energy states with the wave vectors of $\vec{k}+\vec{b}_{n,m}$, where $\vec{b}_{n,m} = n \vec{b}_1 + m \vec{b}_2$ and $n$, $m$ are integers. These potentials are given by the following expressions\cite{PRL_2023_FCI_MacDonald}
\begin{eqnarray}
\Delta_b = & & V \left( e^{i \vec{B}_1\vec{r}} e^{i\psi} +  e^{-i \vec{B}_1\vec{r}} e^{-i\psi}  
+ \right. \nonumber \\ 
& & e^{i \vec{B}_3\vec{r}} e^{i\psi} +
e^{-i \vec{B}_3\vec{r}} e^{-i\psi} + \nonumber \\
& & \left. e^{i \vec{B}_5\vec{r}} e^{i\psi} +
e^{-i \vec{B}_5\vec{r}} e^{-i\psi}
\right)
\end{eqnarray}
\begin{eqnarray}
\Delta_t = & & V\left( e^{i \vec{B}_1\vec{r}} e^{-i\psi} +  e^{-i \vec{B}_1\vec{r}} e^{i\psi}  
+  \right. \nonumber \\ 
& & e^{i \vec{B}_3\vec{r}} e^{-i\psi} +
e^{-i \vec{B}_3\vec{r}} e^{i\psi} + \nonumber \\
& & \left. e^{i \vec{B}_5\vec{r}} e^{-i\psi} +
e^{-i \vec{B}_5\vec{r}} e^{i\psi} 
\right) ,
\end{eqnarray}
\begin{equation}
\Delta_T = w \left(1+ e^{i \vec{B}_5\vec{r}} +
e^{-i \vec{B}_3\vec{r}}   \right) ,
\end{equation}
where $\vec{B}_1 =\vec{b}_1+\vec{b}_2 = b_0 \left( 1, 0 \right)$, $\vec{B}_2 =\vec{b}_2$,
$\vec{B}_3 =-\vec{b}_1$, $\vec{B}_5 =-\vec{b}_2$. As a consequence, the low-energy effective model of the valence band states is characterized by three main parameters: $V$, $w$, and $\psi$. These parameters are obtained from ab initio calculations, which combine the structural relaxation of twisted bilayer system and DFT-type calculations of single-particle energy spectra. Using different methods of ab intio calculations different sets of parameters for the potentials  $\Delta_b$, $\Delta_t$, and $\Delta_T$ were reported by various authors\cite{PRL_2023_FCI_MacDonald,
FCI_Reddy_PRB_2023,FCI_PRL_2024}. 

Given the values of \(V\), \(w\), and \(\psi \), the electron energy spectra are obtained by considering a finite number of higher-energy states in reciprocal space. Specifically, for each \(\vec{k}\)-point in the reciprocal unit cell, we include the coupling of this state to the 22 nearest states, \(\vec{k}\pm \vec{B}_{i}\), which are determined by the coupling terms in \(\Delta _{T}\), \(\Delta _{b}\), and \(\Delta _{t}\). These couplings yield 23 single-particle bands. The wavefunction for each band consists of 23 components and is denoted as \(\Psi _{\mu ,\vec{k},p}\), where \(\mu =1,\ldots, 23\) is the band index and \(p=1,\ldots, 23\) labels the wavefunction component corresponding to a vector of the form \(\vec{k}\pm \vec{B}_{i}\).

In the analysis of electron-electron interaction effects, we consider only a single band, which is nearly flat. In our calculations, this corresponds to the second-highest valence band. We restrict our focus to the states within this band and project all types of interactions considered in this work onto it. To achieve this, we consider a finite-sized electron system and a finite number of single-particle states in the unit reciprocal space. 

In what follows, we consider a finite-size system of \(N_e = 5\) electrons. To describe the \(\nu = 1/3\) incompressible state, the number of points in reciprocal space should be \(N_0 = 15\). Such a number $N_0$ of allowed vectors in the reciprocal space is obtained by placing the system on a torus with periodic boundary conditions that are determined by two vectors $\vec{L}_1$ and $\vec{L}_2$. Then the number $N_0$ is the number of moir\"e unit cells in the supercell formed by vectors $\vec{L}_1$ and $\vec{L}_2$. 
For a FCI with one quasihole-like excitation, the number of such unit cells is \(N_1 = 16\), while for a FCI with one quasiparticle-like excitation, it is \(N_{-1} = 14\). In general, the locations of the mesh points can be defined in terms of two reciprocal lattice vectors, \(\vec{S}_{1}\) and \(\vec{S}_{2}\), corresponding to $\vec{L}_1$ and $\vec{L}_2$, such that the mesh points are given by
\begin{equation}
\vec{k}_{n,m}=n\vec{S}_{1}+m\vec{S}_{2},
\end{equation}
where \(n\) and \(m\) are integers, chosen such that \(\vec{k}_{n,m}\) lies within the reciprocal unit cell, and the total number of points is \(N_{i}\) (where \(i = -1, 0, 1\)).

For a five-electron FCI state with \(N_0 = 15\) single-particle states, the vectors \(\vec{S}_{1}\) and \(\vec{S}_{2}\) are defined as:\( \vec{S}_1 = \frac{\vec{b}_1 + \vec{b}_2}{3}\) and \(\vec{S}_2 = \frac{\vec{b}_1}{5}\). Meanwhile, for the corresponding quasihole and quasiparticle states, these vectors are defined respectively as  \(\vec{S}_1 = \frac{1}{4} \vec{b}_1\), \(\vec{S}_2 = \frac{1}{4} \vec{b}_2\) (\(N_1 = 16\)) and  \(\vec{S}_1 = \frac{1}{2} \vec{b}_1\) and \(\vec{S}_2 = \frac{1}{7} \vec{b}_2\) (\(N_{-1} = 14\)).


Our theoretical approach to determine the physics of quasiparticle and quasihole excitations is to consider a system where the twisted bilayer system is Coulomb coupled to a planar, parabolic quantum dot (QD)\cite{QD_FQHE_my_2002,
QD_FQHE_PRL_2003,spin_transitions_my_PRB_2006,apalkov2026pfaffian}. Most importantly, we found that  the presence of the dot changed locally the spin polarization of
the electron liquid that depends crucially on the separation distance\cite{QD_FQHE_my_2002,
QD_FQHE_PRL_2003,spin_transitions_my_PRB_2006}. Both the bilayer system and the QD\cite{Chakraborty_QDs,
Maksym_QDs_magnetic_field_1990,
Fock_Darwin_graphene_2007} are occupied by electrons. For a finite-size system, the corresponding Coulomb interaction Hamiltonian consists of two terms
\begin{equation}
{\cal H}_{\mathrm{int}} =  
{\cal H}_{\mathrm{b-b}} + {\cal H}_{\mathrm{QD-b}}  
\end{equation}
where \(\mathcal{H}_{\mathrm{b-b}}\) is the electron-electron interaction within the bilayer, and \(\mathcal{H}_{\mathrm{QD-b}}\) is the electron-electron interaction between the QD layer and the bilayer. Both terms represent Coulomb-type interactions. 

For the electron-electron interactions within the bilayer system, the matrix elements between the single-particle states are given by the following expression
\begin{eqnarray}
\left\langle \vec{k}_1, \vec{k}_2 
\right| {\cal H}_{\mathrm{b-b}} \left| 
\vec{k}_3, \vec{k}_4 \right\rangle & = & 
\frac{1}{A_0} \sum_{\vec{b}_i } \sum_{\vec{Q}_i}
\sum_{\vec{Q}_j}
V_{bb} (\vec{q} + \vec{b}_i)  \nonumber \\
& & \times \Psi_{\mu_0, \vec{k}_3, \vec{Q}_i + \vec{b}_i}^*
\Psi_{\mu_0, \vec{k}_4, \vec{Q}_j - \vec{b}_i}^*
\nonumber \\
 & & \times
\Psi_{\mu_0, \vec{k}_1, \vec{Q}_i}
\Psi_{\mu_0, \vec{k}_2, \vec{Q}_j},
\end{eqnarray}
where $\vec{k}_1+\vec{k}_2 = \vec{k}_3+\vec{k}_4$, $\vec{q} = \vec{k}_3-\vec{k}_1$, and all vectors $\vec{b}_i$, $\vec{Q}_i$, and $\vec{Q}_j$ are reciprocal lattice vectors. All summations satisfy the condition that the points $\vec{Q}_i$, $\vec{Q}_j$, $\vec{Q}_i - \vec{b}_i$, and $\vec{Q}_j-\vec{b}_i$ lie within the set of 23 basis reciprocal lattice vectors. The area \(A_{0}\) in the expression for the matrix element is the area of a unit cell of the direct superlattice corresponding to a given set of chosen points, i.e., \(A_0 = \vert{}\vec{L}_1\times \vec{L}_2\vert{}\), where \(\vec{L}_{1}\) and \(\vec{L}_{2}\) are the basis vectors of the direct superlattice corresponding to the reciprocal vectors \(\vec{b}_{1}\) and \(\vec{b}_{2}\).
The Fourier component of the inter-electron interaction $V_{bb}(\vec{q})$ in the bilayer system is\cite{FCI_Reddy_PRB_2023} 
\begin{equation}
V_{bb}(\vec{q}) = \frac{e^2}{4\pi \epsilon q}.
\end{equation}

We assume that the QD is defined by an isotropic parabolic potential with a harmonic oscillator length \(l_{0}\)\cite{Chakraborty_QDs,
Maksym_QDs_magnetic_field_1990}. The corresponding electronic states are labeled by two integers, \(n_1 = 0, 1, \dots\) and \(n_2 = 0, 1, \dots\), with energies given by \(E_{n_1,n_2} = \hbar \omega_0 (n_1+n_2+1)\). Below, we consider only the few lowest-energy QD states: \((n_1,n_2) = (0,0), (1,0), (0,1), (2,0), (0,2), \text{ and } (1,1)\). We relabel these states using a single index, \(N_Q = 1, \dots, 6\). Then, the matrix elements of the interaction Hamiltonian \(\mathcal{H}_{\mathrm{QD-b}}\) are given by the following expression
\begin{eqnarray}
& & \left\langle \vec{k}_1, N_{Q,1} 
\right| {\cal H}_{\mathrm{QD-b}} \left| 
\vec{k}_3, N_{Q_2} \right\rangle = 
\frac{e^2}{\epsilon } \sum_{\vec{Q}_i}
\sum_{\vec{Q}_j}  \int d\vec{r}_{QD} \int d\vec{r}_b \nonumber \\
& &  
\times \frac{1}{\sqrt{(\vec{r}_{QD}-\vec{r}_b)^2 + d_0^2}}  \Psi_{QD,N_{Q,j}}(\vec{r}_{QD}) 
\Psi_{QD,N_{Q,i}}(\vec{r}_{QD})
 \nonumber \\
& & \times \Psi_{\mu_0, \vec{k}_3, \vec{Q}_i + \vec{b}_i}^*
\Psi_{\mu_0, \vec{k}_1, \vec{Q}_i} \exp\left(i\left[   
\vec{k}_1+ \vec{Q}_i -\vec{k}_3- \vec{Q}_j
\right] \vec{r}_b  \right),
\end{eqnarray}
where $d_0$ is the separation between the bilayer and the QD layer. 
Because the QD lacks translational symmetry while the bilayer lacks rotational symmetry, the matrix elements are nonzero between all bilayer and QD states.


{\it Ground state and quasiparticle energies:} We consider a five-electron system within a TMDC bilayer and a single electron in the QD. We address the problem of spin polarization by comparing the energy of a spin-polarized state in the bilayer to a state with one inverted spin. We study the bilayer system for the \(\nu = 1/3\) state, as well as states with one quasihole and one quasielectron. To address how sensitive the results are to the parameters of the bilayer system, we consider three sets of parameters of the low-energy effective model corresponding to MoTe$_2$ twisted bilayer with Hamiltonian (\ref{Ham0}). The parameters are presented in Table \ref{tab:parameters}.

\begin{table}[htbp]
\centering
\caption{Parameters $V$, $w$, and $\psi$ of a single particle low energy Hamiltonian (\ref{Ham0}). These sets of parameters are shown with their corresponding references. }
\label{tab:parameters}
\begin{tabular}{cccc}
\toprule
Parameter & \textbf{Set 1}\cite{PRL_2023_FCI_MacDonald}
 & \textbf{Set 2}\cite{FCI_Reddy_PRB_2023} 
 & \textbf{Set 3}\cite{FCI_PRL_2024}  \\
\hline
$V$ & 8.0 meV & 11.2 meV & 20.8 meV  \\
$w$ & -8.5 meV & -13.3 meV & -23.8 meV  \\
$\psi$ & -89.6$^\circ $ & -91.0$^{\circ}$ & -107.7$^{\circ}$ \\
\hline
\end{tabular}
\end{table}

For a finite-size system, we use the exact diagonalization approach to find the ground-state energy. First, we find the energy \(E_{s-p}\) of the ground state for the spin-polarized bilayer system—that is, when all five electrons have the same spin-up direction. Then, we find the ground-state energy \(E_{s-u}\) of the spin-unpolarized twisted system, where four electrons have a spin-up direction and one electron has a spin-down direction. We then characterize the energetics of the system's spin polarization using the energy difference $\Delta E = E_{s-p} - E_{s-u}$. We evaluate the \(\Delta E\) both for a pure twisted TMDC bilayer without being coupled to a QD and for a TMDC bilayer coupled to a QD that contains a single electron. The results of these calculations are summarized in Tables \ref{tab:FCI},\ref{tab:FCI_QH}.

The results for the \(\nu = 1/3\)-FCI state are shown in Table \ref{tab:FCI}. For a pure bilayer system without any external perturbation from the QD, the ground state is spin-polarized (i.e., \(\Delta E < 0\)), where the energy difference between the spin-polarized and spin-unpolarized states ranges from \(4\) to \(10\) meV, depending on the parameter set. When the Coulomb coupling to the QD is introduced, the ground state becomes spin-unpolarized (i.e., \(\Delta E > 0\)). For Sets 2 and 3, the energy gap \(\Delta E\) is approximately \(10\) meV, whereas for Set 1, it is slightly smaller, at around \(5\) to \(8\) meV. This energy gap depends on the separation between the bilayer and the QD, showing a monotonic decrease with \(d_{0}\) for Sets 1 and 3. Therefore, the perturbation of the FCI state by a QD results in the formation of a spin-unpolarized state. The interaction of the bilayer system in an incompressible FCI state with an electron in a QD excites quasihole-quasiparticle pairs. The results presented in Table \ref{tab:FCI} suggest that such excitations are spin-unpolarized. To address the spin polarization properties of the elementary excitations of an FCI, we consider a FCI system containing one quasihole or one quasiparticle.

\begin{table}[h]
\centering
\caption{$\nu =1/3$-FCI state. The energy difference \(\Delta E\) is shown for three sets of parameters and for five different systems: the pure bilayer system (second column) and the bilayer system coupled to a QD system for four different separations \(d_{0}\) between the bilayer system and the QD layer, where the distance \(d_{0}\) is given in nm. The energy difference $\Delta E$ is given in meV. The bilayer system consists of five electrons that are in $\frac{1}{3}$-FCI state, i.e., the number of single particle state in bilayer is 15. The QD is occupied by only one electron.   }
\label{tab:FCI}
\begin{tabular}{ccccccc}
\hline
\textbf{Set} $\vert$ & bilayer $\vert$ & 
$d_0=1.0$ $\vert$ & $d_0=1.5$ $\vert$ & 
$d_0=2.0$ $\vert$ & $d_0=2.5$ $\vert$ & $d_0=3.0$ \\
\hline
1  & -4.297  & 8.051     & 7.532  & 6.980 & 6.387 & 5.783 \\
2  & -10.113  & 10.471 & 10.461   & 10.481 & 10.531 & 10.613 \\
3  & -4.983  & 10.656  & 10.386 & 10.222   & 10.141 & 10.134  \\ \hline
\end{tabular}
\end{table}

Table \ref{tab:FCI_QH} presents the data for the FCI with one quasihole. Without a coupling to the QD system, the FCI quasihole is spin-polarized, with a corresponding spin-flip energy gap of around 5 meV that depends weakly on the bilayer system parameters. When the bilayer system is coupled to a QD, the quasihole state becomes spin-unpolarized. The corresponding spin-flip energy gap is approximately 10 meV, reaching a maximum value of around 12 meV for Set 3. As the separation between the bilayer and the QD increases, the energy gap monotonically increases for \(d_{0}\) up to 3 nm, beyond which the dependence on \(d_{0}\) is weak. Across the different sets, the energy gaps are similar, with the highest value observed for Set 3. These energy gaps are slightly higher—by roughly 1 meV—than the corresponding energy gaps for the FCI state shown in Table \ref{tab:FCI}.

\begin{table}[h]
\centering
\caption{$\nu =1/3$-FCI state with one quasihole. The energy difference \(\Delta E\) is shown for three sets of parameters and for five different systems: the pure bilayer system (second column) and the bilayer system coupled to a QD system for four different separations \(d_{0}\) between the bilayer system and the QD layer, where the distance \(d_{0}\) is given in nm. The energy difference $\Delta E$ is given in meV. The bilayer system consists of five electrons that are in $\frac{1}{3}$-FCI state with one quasihole, i.e., the number of single particle state in bilayer is 16. The QD is occupied by one electron. }
\label{tab:FCI_QH}
\begin{tabular}{ccccccc}
\hline
\textbf{Set} $\vert$ & bilayer $\vert$ & 
$d_0=1.0$ $\vert$ & $d_0=1.5$ $\vert$ & 
$d_0=2.0$ $\vert$ & $d_0=2.5$ $\vert$ & $d_0=3.0$ \\
\hline
1 & -5.703 & 9.947 & 10.286 & 10.705 & 11.199 & 11.768 \\
2 & -4.973 & 9.690 & 10.158 & 10.733 & 11.368 & 12.033 \\
3 & -3.535 & 12.162 & 12.343 & 12.570 & 12.831 & 13.116 \\
\hline
\end{tabular}
\end{table}

A larger energy scale is observed for the quasiparticle excitations. Table \ref{tab:FCI_QP_negative} shows the results for the FCI with one quasiparticle. For the uncoupled bilayer-QD system, the quasiparticle state is spin-polarized with a large spin-flip gap: 11 meV for Set 2 and 18 meV for Set 1. However, for parameter Set 3, the energy gap is much smaller, at around 2 meV. Notably, for the pure FCI and the FCI with one quasihole, the spin-flip energy gaps across all parameter sets are comparable.

The coupling of the FCI quasiparticle to the QD favors a spin-flipped ground state. The corresponding energy gaps are relatively large, at approximately 25–30 meV for the parameter sets 1 and 2. This is two to three times larger than the corresponding values for the FCI quasihole state (see Table \ref{tab:FCI_QH}). For parameter Set 3, the spin-flip energy gap is around 7 meV, which is much smaller than the values for sets 1 and 2; this behavior also differs from the results obtained for the FCI quasihole state. Therefore, the spin energetics of the quasiparticle state exhibit a strong dependence on the specific parameter set. 

The results shown in Table \ref{tab:FCI_QP_negative} correspond to a QD occupied by a single electron. In this case, an extra, negatively charged FCI quasiparticle will be repelled by the QD, although this effect is not clearly visible in our small-sized bilayer system. To further probe the properties of the FCI coupled to the QD, we also analyze a QD occupied by a single hole, which is positively charged. In this scenario, the FCI quasiparticle is expected to form a bound state with the QD. The corresponding results are presented in Table \ref{tab:FCI_QP_positive}. Similar to the case of the negatively charged QD (see Table \ref{tab:FCI_QP_negative}), the FCI quasiparticle state becomes spin-unpolarized. However, the spin-flip energy gaps for the positively charged QD are approximately 5 meV larger than those for the negatively charged QD. This suggests that the quasiparticle is located closer to the QD, forming a bound QD–quasiparticle state that exerts a stronger effect on the spin energetics of the system. 

The spin-flip energy gaps for the pure FCI and the FCI with a quasihole are comparable, and both are much smaller than the corresponding energy gaps for the FCI with a quasiparticle system. This suggests that in a FCI bilayer coupled to a QD, a spin-polarized quasiparticle and a spin-unpolarized quasihole are created.

\begin{table}[h]
\centering
\caption{$\nu =1/3$-FCI state with one quasiparticle. The energy difference \(\Delta E\) is shown for three sets of parameters and for five different systems: the pure bilayer system (second column) and the bilayer system coupled to a QD system for four different separations \(d_{0}\) between the bilayer system and the QD layer, where the distance \(d_{0}\) is given in nm. The energy difference $\Delta E$ is given in meV. The bilayer system consists of five electrons that are in $\frac{1}{3}$-FCI state with one quasiparticle, i.e., the number of single particle state in bilayer is 14. The QD is occupied by one electron. }
\label{tab:FCI_QP_negative}
\begin{tabular}{ccccccc}
\hline
\textbf{Set} $\vert$ &  bilayer $\vert$ & 
$d_0=1.0$ $\vert$ & $d_0=1.5$ $\vert$ & 
$d_0=2.0$ $\vert$ & $d_0=2.5$ $\vert$ & $d_0=3.0$ \\
\hline
1 & -18.424 & 27.640 & 27.865 & 28.109 & 28.362 & 28.614 \\
    2 & -11.077 & 23.736 & 23.821 & 23.917 & 24.018 & 24.108 \\
    3 & -1.844  &  7.880 &  7.769 &  7.657 &  7.558 &  7.480 \\
\hline
\end{tabular}
\end{table}

\begin{table}[h]
\centering
\caption{$\nu =1/3$-FCI state with one quasiparticle. The energy difference \(\Delta E\) is shown for three sets of parameters and for five different systems: the pure bilayer system (second column) and the bilayer system coupled to a QD system for four different separations \(d_{0}\) between the bilayer system and the QD layer, where the distance \(d_{0}\) is given in nm. The energy difference $\Delta E$ is given in meV. The bilayer system consists of five electrons that are in $\frac{1}{3}$-FCI state with one quasiparticle, i.e., the number of single particle state in bilayer is 14. The QD is occupied by one hole, i.e., positively charged particle. }
\label{tab:FCI_QP_positive}
\begin{tabular}{ccccccc}
\hline
\textbf{Set} $\vert$ & bilayer $\vert$ & 
$d_0=1.0$ $\vert$ & $d_0=1.5$ $\vert$ & 
$d_0=2.0$ $\vert$ & $d_0=2.5$ $\vert$ & $d_0=3.0$ \\
\hline
1 & -18.424 & 31.740 & 31.687 & 31.632 & 31.577 & 31.526 \\
    2 & -11.077 & 26.968 & 26.961 & 26.951 & 26.943 & 26.934 \\
    3 & -1.844  & 11.484 & 11.331 & 11.178 & 11.018 & 10.858 \\
\hline
\end{tabular}
\end{table}

\begin{figure}[b]
\includegraphics[width=0.8\columnwidth]{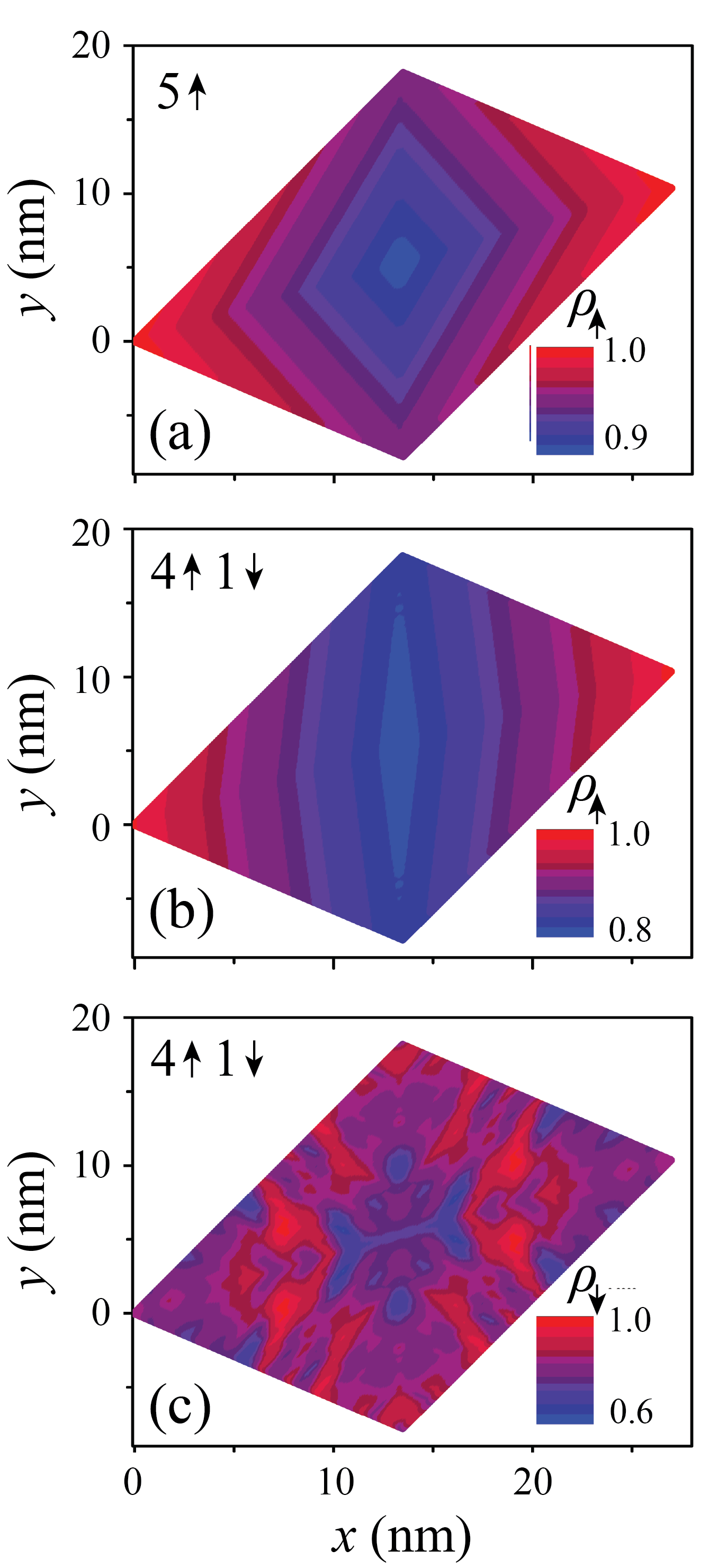}
\caption{\label{FCI_all}  
Electron density for a $\frac{1}{3}$-FCI state coupled to a QD, shown within one supercell. (a) Spin-up electron density $\rho _{\uparrow }$ for a bilayer system containing five spin-up electrons. (b, c) Charge densities for a system with four spin-up electrons and one spin-down electron, showing (b) $\rho _{\uparrow }$ and (c) $\rho _{\downarrow }$. All densities are normalized to their maximum values. The distance between the bilayer system and the QD layer is $d_0 = 2$ nm. 
 } 
\end{figure}

\begin{figure}[b]
\includegraphics[width=0.8\columnwidth]{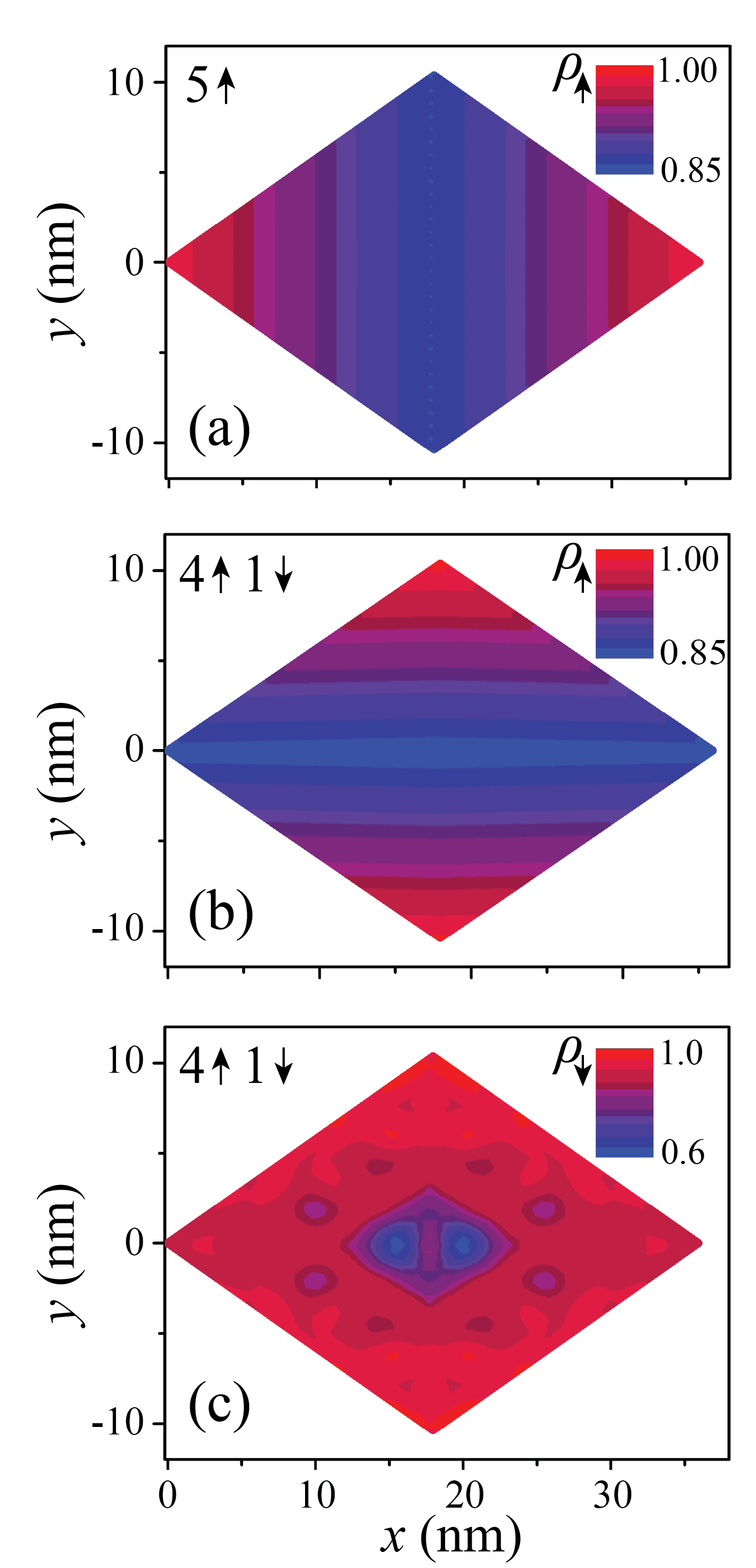}
\caption{\label{QHOLE_all}  
Electron density for a $\frac{1}{3}$-FCI state with one quasihole coupled to a QD, shown within one supercell. (a) Spin-up electron density $\rho _{\uparrow }$ for a bilayer system containing five spin-up electrons. (b, c) Charge densities for a system with four spin-up electrons and one spin-down electron, showing (b) $\rho _{\uparrow }$ and (c) $\rho _{\downarrow }$. All densities are normalized to their maximum values. The distance between the bilayer system and the QD layer is $d_0 = 2$ nm.
 } 
\end{figure}

\begin{figure}[b]
\includegraphics[width=0.8\columnwidth]{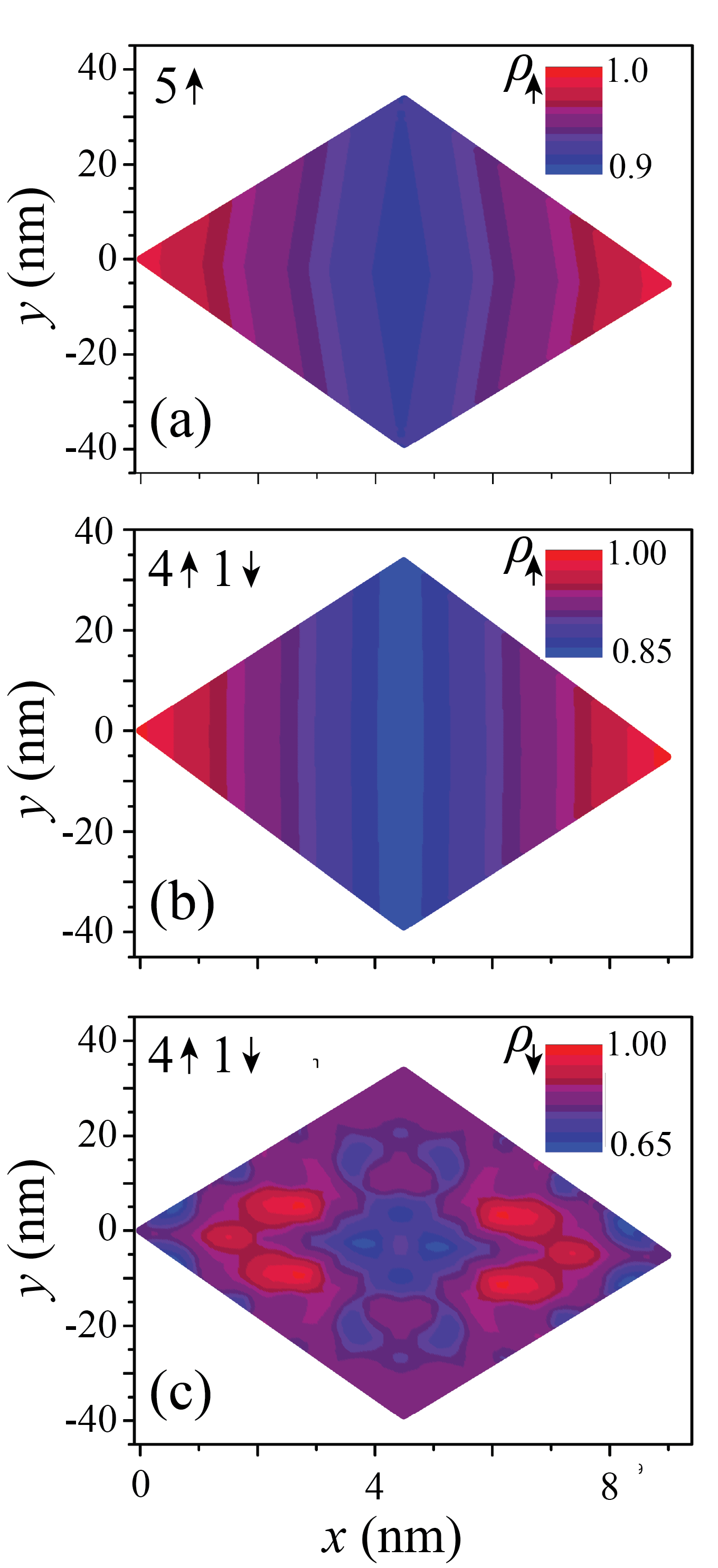}
\caption{\label{QPARTICLE_all}  
Electron density for a $\frac{1}{3}$-FCI state with one quasiparticle coupled to a QD, shown within one supercell. (a) Spin-up electron density $\rho _{\uparrow }$ for a bilayer system containing five spin-up electrons. (b, c) Charge densities for a system with four spin-up electrons and one spin-down electron, showing (b) $\rho _{\uparrow }$ and (c) $\rho _{\downarrow }$. All densities are normalized to their maximum values. The distance between the bilayer system and the QD layer is $d_0 = 2$ nm.
 } 
\end{figure}

{\it Charge density:} One of the characteristics of the correlated many-particle state is the charge density. We define the charge density for electrons with spin-up and spin-down components from the following expression 
\begin{equation}
\rho_{s} (\vec{r}) = \left\langle \Phi\right| \sum_{i} \delta(\vec{r} - \vec{r}_{i,s}) \left| \Phi \right\rangle ,
\end{equation}
where $\left| \Phi\right\rangle$ is the ground state wavefunction of many-particle system, $s=\uparrow$ or $\downarrow$ is the spin component, and 
$\vec{r}_{i,s}$ is the coordinate of an electron with spin $s$.
We calculate the charge density within the superlattice that is determined by vectors $\vec{L}_1$ and $\vec{L}_2$. The results are shown in Figs.~\ref{FCI_all}, \ref{QHOLE_all}, and \ref{QPARTICLE_all}. For the $\frac{1}{3}$-FCI state in the twisted bilayer [Fig.~\ref{FCI_all}(a)], where the QD excites quasihole-quasiparticle pairs, the charge density profile shows a clear localization of the excited quasihole near the QD, which is placed at the center of the supercell. When one spin is flipped [Figs.~\ref{FCI_all}(b) and (c)], the charge profile shows a strong localization of the positive charge near the QD for the spin-down component and a more delocalized profile for the spin-up component [Fig.~\ref{FCI_all}(b)].

For the twisted bilayer systems with one quasihole [Fig.~\ref{QHOLE_all}] or one quasiparticle [Fig.~\ref{QPARTICLE_all}], the situation differs from that of the pure $\frac{1}{3}$ state [Fig.~\ref{FCI_all}]. Specifically, for the spin-polarized system [Fig.~\ref{QHOLE_all}(a) and \ref{QPARTICLE_all}(a)], the charge density shows a localized accumulation of positive charge along only one direction, while it remains largely delocalized along the other direction. Similar delocalization features are visible in the density profiles for the spin-up component in the spin-unpolarized case [Figs.~\ref{QHOLE_all}(b) and \ref{QPARTICLE_all}(b)]. At the same time, the density profiles for the spin-down component show a strong localization of the positive charge near the QD [Figs.~\ref{QHOLE_all}(c) and \ref{QPARTICLE_all}(c)]. This behavior correlates with the energetics of the spin-polarized and spin-unpolarized systems, where the spin-unpolarized system exhibits a lower energy due to the strong accumulation of positive charge near the QD within the spin-down component.

{\it Conclusion:} Theoretical prediction of the zero-field FQHE in fractional Chern insulators and its 
subsequent experimental confirmation stand as a true scientific tour de force. Here 
we have explored the properties of Laughlin-type quasiparticles and quasiholes, not least their
spin properties in a twisted bilayer MoTe$_2$ system in close proximity to a QD containing
a single electron or a hole.  We found that the incompressible bilayer FCI state, when interacts
with the QD, excites spin-unpolarized quasiparticle-quasihole pairs. In the case of the quasiholes,
the spin-flip energy gap has a slight dependence on the separation distance
with the QD. A larger energy gap is observed for quasiparticle excitations. In the case of the QD 
containing a positively-charged hole the quasiparticle state is also
spin unpolarized. The quasiparticle is found to be located closer to the QD forming a 
bound QD-quasiparticle state. The charge density profile shows a strong localization for the spin-flipped component, which correlates with the fact that the spin-unpolarized state has a lower energy compared to the spin-polarized one. Thus, the spin-reversed quasiparticles are characterized by a strong accumulation of positive charge near the negatively charged QD for the spin-reversed component.

Compared to its finite-magnetic field counterpart, the zero-field FQHE in the Moir\'e bilayer lattice suggests
a strong presence of spin-reversed qasiholes and quasiparticles that could potentially open up the
door to explore exciting spin-related phenomena in the FQHE, such as zero-field fractional skyrmions 
in a Laughlin state\cite{FQHE_measurements_PRL_1997}, or the Laughlin spin-liquid states\cite{Laughlin_state_PRL_2012}. Unlike in conventional FQHE, where
the spin properties are explored with a tilted magnetic 
filed\cite{FQHE_tilted_field,Subband_tilted_PRB1990,
Spin_configurations_FQHE,Evidence_phase_transition_FQHE}, in the FCI in a Moir\'e bilayer
system, the photoluminescence spectroscopy is perhaps is the best option\cite{QD_FQHE_PRL_2003,
Signature_fractional_nature_2026,anyons_bosonic_properties_2021} to observe the
spin effects predicted here.

The funding was provided by by Grant No. DE-SC0007043
from the Materials Sciences and Engineering Division of
the Office of the Basic Energy Sciences, Office of Science,
US Department of Energy.


%

\end{document}